\documentclass[10pt,preprintnumbers,twocolumn,nofootinbib,floatfix]{revtex4}

\usepackage{epsfig}
\usepackage{amssymb}
\usepackage{amsmath}
\usepackage{mathrsfs}

\makeindex \textheight=9.25in \textwidth=6.5in \topmargin=-0.25in
\begin{document}
\title{Unfolding Geometric Unification in M-Theory}
\author{Jacob L. Bourjaily}
\email{jbourjai@princeton.edu}
\affiliation{Joseph Henry Laboratories, Princeton University, Princeton, NJ 08544}
\date{22$^{\mathrm{nd}}$ June 2007}

\begin{abstract}
By reinterpreting the familiar tools and ideas of M-theory model building, we show how a $G_2$-manifold locally engineered to give rise to massless matter representations of an $SU_5$ grand unified model can be smoothly unfolded into a $G_2$-manifold giving rise to $SU_3\times SU_2$ gauge theory with the corresponding matter representations. These ideas could lead to new insights in string phenomenology because much of the arbitrariness of M-theory model building can be removed by supposing, for example, that the singularities giving rise to Standard Model particles could arise from unfolding a more singular, grand unified geometry. 
\end{abstract}

\maketitle
\section{Introduction}
Massless matter charged under a non-Abelian gauge group is known to arise in M-theory compactified on $G_2$ in two ways, both of which are geometric in origin: by the existence of boundaries in the compactification manifold (e.g. Ho\v{r}ava-Witten theory \cite{Horava:1996ma}) or by the existence of isolated (co-dimension seven) enhancements of co-dimension four singularities \cite{Atiyah:2001qf,Witten:2001uq,Acharya:2001gy,Acharya:2004qe}. In this paper, we will be interested in the latter. The description of how singularities in the compactification manifold can give rise to gauge theory and massless matter representations is known as geometrical engineering and has applications to type IIa string theory and F-theory as well \cite{Katz:1996xe}. 

The singular geometries giving rise to most desirable phenomenological matter representations are well known in the literature (see e.g \cite{Acharya:2001gy,Berglund:2002hw}). Until recently, however, it was hard to see how the framework of geometrical engineering itself could be used to motivate phenomenological models. The reason is that although one can describe the local geometry giving rise to almost any desirable matter representation, a phenomenological model would seem to require dozens of these isolated singularities. Without duality to a concrete string construction\footnote{Geometrically engineered M-theory constructions are dual to intersecting D-brane models in type IIa, for example.}, there appeared to be little to motivate the numbers, types, and relative positions of all the singularities required. Indeed, simply engineering a manifold to give rise to $SU_3\times SU_2$ gauge theory with one enhanced singularity for every known (and anticipated) particle of the Standard Model would seem to be wishful thinking at best, and arbitrary worst---why only those singularities? and why at those relative positions?

A possible way to motivate the many singularities required to engineer the Standard Model could be a geometric analogue of grand unification \cite{Bourjaily:2007vw,Bourjaily:2007vx}. In the context of type IIa string theory, we showed in \cite{Bourjaily:2007vx} how a geometrically-engineered $\mathscr{N}=2$ model with $E_6\times SU_2$ gauge theory and a single $E_8$-type conical singularity could be deformed into a geometrically-engineered model with $SU_3\times SU_2$ gauge theory with 35 isolated conical singularities of the types required to reproduce three families. The numbers and types of singularities of the `unfolded' model---and in particular that three families emerges---is a consequence of group theory and algebraic geometry alone. And their relative locations, which determine the superpotential, are set by the values of the moduli which deform the initial, maximally singular geometry.

This then is the motivation underlying our present discussion: the possibility that the many disparate singularities required for the Standard Model matter fields could arise from the `unfolding' of a manifold engineered to give rise to some level of grand unification. In \cite{Bourjaily:2007vw,Bourjaily:2007vx} we showed how this could be realized concretely in the context of type IIa string theory (and to some extent in F-theory). But a concrete demonstration that this story could be realized in M-theory was lacking. In this paper, we fill in this gap by showing explicitly how local, $G_2$-manifolds engineered to give rise to $SU_5$ representations in M-theory can be smoothly deformed into those with multiple, isolated conical singularities giving rise to $SU_3\times SU_2$ matter representations. 

The basic ingredients and necessary tools are well known in the literature and the story we present here is in many ways a natural extension of traditional geometrical engineering. What is new in our present work is the application of these well-known tools used to describe the local geometry about a {\it single} conical singularity of a given type to now also describe the local geometry about an {\it entire collection of disparate conical singularities} which arise by the unfolding of a more singular (more unified) initial geometry. And this is done by a geometric analogue to unified symmetry breaking. 

There are several motivations for studying these `unfolded geometries' beyond those of natural esthetics. First, the framework allows us to describe concretely the total geometry about most or all of the phenomenologically relevant structures in the manifold. And the local structure that emerges can be very constrained: all of the relative positions of disparate singularities are specified in terms or a small number of deformation moduli---and this greatly reduces the arbitrariness of the form of the superpotential, for example: insisting on a local, unfolded perspective puts many restrictions on the range of possible models. And not unimportantly, the locality of these constructions allows us to study possible phenomenological predictions of M-theory before solving all of quantum gravity, much in the spirit of \cite{Verlinde:2005jr}. 

We will limit ourselves in this paper to the example of unfolding matter representations of $SU_5$ into those of $SU_3\times SU_2$. This is done both as a proof of concept of the over-arching idea, and to serve as a pedagogical example of the methods and tools involved. It turns out that unfolding $SU_5$ representations in M-theory is just simple enough---but barely so---that we can guess the necessary deformed geometry and verify our guesses by inspection. A more sophisticated approach would be required to systematically study the unfolding of more unified singularities. In particular, this means that the more phenomenologically appealing example of unfolding three families out of an isolated $E_8$-type singularity (as done in \cite{Bourjaily:2007vx} in type IIa and F-theory) must wait for a future work.

In section \ref{geoengin}, we review the basic ingredients of geometrical engineering in M-theory and describe the local geometries which give rise to the $\mathbf{5}$ and $\mathbf{10}$ representations of $SU_5$. In section \ref{unfold} we describe how to deform these such that the resulting geometry gives rise to $SU_3\times SU_2$ gauge theory; when this is done, we find that the single singularities giving rise to $SU_5$ representations are `broken apart' into separated singularities producing their corresponding Standard Model particle content.

\section{Geometrical Engineering in M-theory\label{geoengin}}
Geometrical engineering in M-theory is qualitatively the same as that in type IIa. In both, gauge theory arises via co-dimension four singular surfaces in the compactification manifold, and charged matter arises if there are {\it isolated} points on these surfaces where the singularity is enhanced. The way in which the singularity is enhanced determines the representation that results. The only difference between engineering in M-theory and type IIa is the dimension of the singular surfaces giving rise to gauge theory: in type IIa they are two-dimensional while in M-theory they are three-dimensional. But this difference does have an important consequence: it makes it possible to differentiate the geometry giving rise to a representation and its complex conjugate. That is, geometrical engineering in M-theory results in manifestly chiral representations \cite{Witten:2001uq,Acharya:2001gy}.

At any point along the three-dimensional surface of one of these singularities, the geometry is locally equivalent to an ADE-type singularity\footnote{ADE singularities are so-named because they come in the types $A_{n-1}(\equiv SU_n)$, $D_n(\equiv SO_{2n})$, and $E_n$.}. If the type of ADE singularity were the same for every point of three-dimensional surface, then the $G_2$-compactification manifold would give rise to pure gauge theory of the corresponding type. Conveniently, ADE singularities are named according to the gauge theory that results.

In order to describe locations where these singularities are enhanced, we should consider our compactification manifold as a fibration of $K3$ surfaces containing ADE singularities over a three-dimensional base, say $Q$, which is locally coordinatized by a real three-vector $\vec{\!{~}\!t}$. There must be isolated points over which the type of fibre is enhanced by one rank. Looking at this from the point of view of the enhanced fibre, what is needed is a three-dimensional space of deformations of an ADE singularity of one type into another of lesser rank.

(In \cite{Bourjaily:2007vw,Bourjaily:2007vx}, we needed the singularities to be enhanced only at isolated points over a two-dimensional base space. By defining the ADE singularities as singular hyper-surfaces in $\mathbb{C}^3$, we were able to use one-complex dimensional complex structure deformations to achieve local enhancements. This language allowed us to make use of some powerful mathematical knowledge about two-dimensional deformations of ADE singularities (see e.g. \cite{Katz:1992ab}). However, this is not the right language in which to discuss the three-dimensional deformations required for M-theory.)

The description of ADE singularities which makes manifest the three-dimensional nature of their resolutions was given by Kronheimer in \cite{Kronheimer:1989zs,Kronheimer:1989pu}. It was this framework that Acharya and Witten used in \cite{Acharya:2001gy} to describe the local geometry of $G_2$-manifolds engineered to give rise to charged matter for a number of example representations. Their work was extended by Berglund and Brandhuber in \cite{Berglund:2002hw} to include more examples, and the suggestion of multiply unfolding these geometries. In many ways, our present paper is a natural extension and application of the ideas expressed in \cite{Acharya:2001gy} and \cite{Berglund:2002hw}.

\newpage

\subsection{ADE Singularities as hyper-K\"{a}hler Quotients}\vspace{-0.35cm}
A prerequisite for describing a $G_2$-manifold as a fibration of ADE singularities over $Q$ is an adequate description of the ADE fibres themselves. We have already pointed out that their description in terms of complex structure alone will not be adequate for our present purposes. Rather, we will use Kronheimer's construction of ADE singularities as hyper-K\"{a}hler quotients \cite{Kronheimer:1989zs,Kronheimer:1989pu}.

\vspace{-0.5cm}
\subsubsection{$SU_n$ Singularities}\vspace{-0.35cm}
$SU_n$ singularities are both the easiest to describe and can also be the starting point to construct $SO_{2n}$ and $E_n$ singularities. An $SU_n(\equiv A_{n-1})$ singularity is locally equivalent to $\mathbb{R}^4/\mathbb{Z}_{n}$, which is locally equivalent to the surface of solutions to $xy=z^n$ in $\mathbb{C}^3$. Rather than these descriptions, however, we choose to view this space as the vacuum manifold of a particular linear sigma model of $n$ hypermultiplets $\Phi_i$, $i=0,1,\ldots,(n-1)$, with gauge group $K\equiv U(1)^{n-1}$ where under the $i^{\mathrm{th}}$ $U(1)$, $\Phi_i$ has charge $+1$, $\Phi_{i-1}$ has charge $-1$, and other hypermultiplets are neutral. The vacuum manifold is obtained by imposing all the Fayet-Iliopoulos D/F-term constraints (setting them to zero) and taking an ordinary quotient by $K$. We will denote the scalars of $\Phi_i$ by $(z_i,\overline{z}_i)$, with $z_i,\overline{z}_i\in\mathbb{C}$, and let $\mathbb{H}$ denote the space spanned by $z_i$ and $\overline{z}_i$\footnote{We call this space $\mathbb{H}$ to highlight its quaternionic structure.}. The vacuum of this linear sigma model will be called $\mathbb{H}^n//K$.

We will now show that $\mathbb{H}^n//K$ is in fact an $SU_n$ singularity. Notice first that the dimension is correct: start with $4n$ dimensions of $\mathbb{H}^n$; then, applying the D/F-term constraints reduces this by $3(n-1)$; and dividing by $K$ reduces by another $(n-1)$, yielding $\mathrm{dim}\left(\mathbb{H}^n//K\right)=4$. To show that the space is equivalent to the surface $xy=z^n$ will require some additional notation that will be used throughout the paper.

First, recall that the Fayet-Iliopoulos D/F-terms are linear combinations of the moment maps \mbox{$\mu:\mathbb{H}\to\mathbb{R}^3$}, which we will write \begin{equation}\mu:(z_i,\overline{z}_i)\mapsto\left(\begin{array}{c}\mathfrak{Re}\left(z_i\overline{z}_i\right)\\\mathfrak{Im}\left(z_i\overline{z}_i\right)\\\left|z_i\right|^2-\left|\overline{z}_i\right|^2\end{array}\right)\equiv\Phi_i^{\dag}\vec{\sigma}\Phi_i.\end{equation}
Notice that the $U(1)$-charges of $z_i$ and $\overline{z}_i$ are opposite. In this notation, and with the $U(1)$-charge assignments described above, the D/F-term for the $i^{\mathrm{th}}$ $U(1)$ can be written \begin{equation}\vec{\!\!~t}_i\equiv\Phi_i^{\dag}\vec{\sigma}\Phi_i-\Phi_{i-1}^{\dag}\vec{\sigma}\Phi_{i-1}.\end{equation}

The vanishing of all the D/F-terms therefore implies the condition that $z_i\overline{z}_i=z_j\overline{z}_j$ for any $i,j$. This suggests that we introduce the following $K$-invariant variables \vspace{-0.05cm}\begin{equation}x\equiv\prod_{i=0}^{n-1}z_i,\quad y\equiv\prod_{i=0}^{n-1}\overline{z}_i,\quad z\equiv z_0\overline{z}_0.\label{xyz}\end{equation} Because these variables are related by $xy=z^n$, it is clear that the space $\mathbb{H}^n//K$ is indeed equivalent to an $SU_n$ singularity.

The three-dimensional deformations of the singularity are then found by allowing a linear combination of the D/F-terms to be non-vanishing.

\vspace{-0.35cm}\subsubsection{$SO_{2n}$ Singularities}\vspace{-0.35cm}
Following the discussion in \cite{Berglund:2002hw}, we will consider the $SO_{2n}$ singularity to be a $\mathbb{Z}_2$-orbifold of an $SU_{2n}$ singularity. That this is the right geometry to consider is made clear by duality to type IIa: recall that geometrically engineered models in M-theory are dual to intersecting D-brane models in type IIa; with this in mind, the geometry in M-theory giving rise to $SO_{2n}$ gauge theory must dimensionally reduce to a theory with $2n$ stacked D-branes together with an $\mathcal{O}$-plane to generate $SO_{2n}$ gauge theory in type IIa; the presence of the $\mathcal{O}$-plane in type IIa suggests that an appropriate $\mathbb{Z}_2$-orbifold of an $SU_{2n}$ singularity will yield the desired geometry.

Let us start with an $SU_{2n}$ singularity as described above: $\mathbb{H}^{2n}//K$. Following \cite{Berglund:2002hw}, let the action of $\mathbb{Z}_2=\langle S\rangle$ be generated by \vspace{-0.05cm}\begin{equation}S:(z_i,\overline{z}_i)\mapsto(\overline{z}_i,-z_i),\vspace{-0.15cm}\end{equation}or, equivalently, using the notation introduced in equation (\ref{xyz}), \vspace{-0.10cm}\begin{equation}S:(x,y,z)\mapsto(y,x,-z).\vspace{-0.15cm}\end{equation}

Clearly $x,y,$ and $z$ are not $S$-invariant variables. However, they can easily be combined into the related quantities \vspace{-0.15cm}\begin{align}
X\equiv\frac{1}{2}z^{-1}\left(x-y\right),\quad Y\equiv\frac{1}{2}z^{-2}\left(x+y\right),\quad Z\equiv z^2,\label{XYZ}\vspace{-0.15cm}\end{align}
which are indeed seen to be $S$-invariant. Furthermore, they are related by the equation \begin{equation}X^2=Y^2Z-Z^{-1}Z^{n},\end{equation}
which is visibly the defining equation of an $SO_{2n}$ singularity. Therefore, $\left[\mathbb{H}^{2n}//K\right]/\mathbb{Z}_2$ is in indeed the desired geometry.

The deformations of an $SO_{2n}$ singularity are those of the covering space $SU_{2n}$ that are $\mathbb{Z}_2$-invariant. \\

\vspace{-0.65cm}\subsection{Engineering Charged Matter in M-Theory}\vspace{-0.35cm}
The above discussion may at first seem unnecessarily cumbersome. The work will pay off, however, when we observe that it naturally allows for the construction of $G_2$-manifolds with the isolated singularities required to give rise to massless charged matter representations. This was an essential insight of Acharya and Witten in \cite{Acharya:2001gy}. The basic idea is as follows. If one of the $U(1)$ factors in $K$ used above were ignored, say $K'=K\setminus U(1)_j$, then $\mathbb{H}^n//K'$ would be a hyper-K\"{a}hler, eight-dimensional manifold. By taking a normal quotient of this by $U(1)_j$ in such a way that $U(1)_j$ commutes with the three complex structures of $\mathbb{H}^n//K'$, the resulting space $\left(\mathbb{H}^n//K'\right)/U(1)_j$ would be a seven-fold, inheriting a $G_2$-structure from the hyper-K\"{a}hler structure\footnote{See e.g. \cite{Acharya:2004qe} for more details.}.

To be more precise, let $Q$ be some three-dimensional base space which is locally coordinatized by $\vec{\!\!~t}_j$, the D/F-term of $U(1)_j$. Clearly any $\vec{\!\!~t}_j=$constant slice through $\left(\mathbb{H}^n//K'\right)/U(1)_j$ will be four-dimensional and have singularities corresponding to those generated by the other $U(1)$ factors. It is easy to appreciate that when $\vec{\!\!~t}_j\to0$ the singularity is simply the original $SU_n$ singularity. However, for non-vanishing $\vec{\!\!~t}_j$, the singularity is softened by one-rank. This can be understood as using $\vec{~\!\!t}_j$ to blow-up one of the nodes of the Dynkin diagram for $SU_n$.

As described in \cite{Acharya:2001gy}, if the $j^{\mathrm{th}}$ $U(1)$ is taken out of $K$ and its D/F-term is used to coordinatize a local patch on $Q$, then $SU_j\times SU_{n-j}$ gauge theory results with chiral matter in the $(\mathbf{j},\overline{\mathbf{n-j}})$ representation, localized at a single point on $Q$---at $\vec{\!\!~t}_j=0$. The geometry can be shown to be a cone over $\mathbb{WCP}^3_{j,j,(n-j),(n-j)}$. We will not need to discuss this level of detail, however, in our examples below.

\vspace{-0.55cm}\subsubsection{Engineering a $\mathbf{5}$ of $SU_5$\label{5ofsu5}}\vspace{-0.35cm}
We have all the tools necessary to describe the engineering of a $\mathbf{5}$ of $SU_5$. Recalling that this representation results from the resolution $SU_6\to SU_5$, it is clear that we would like to start with the hyper-K\"{a}hler quotient description of an $SU_6$ singularity, and allow one of the D/F-terms to coordinatize a region of the base space $Q$. To reiterate both the concepts involved and our notation, recall that the $SU_6$ singularity is viewed as the vacuum manifold $\mathbb{H}^6//K$ of the linear sigma model of $6$ hyperrmultiplets $\Phi_i$, $i=0,1,\ldots,5$---whose scalars are denoted $(z_i,\overline{z}_i)$---that are charged under $K\equiv\prod_{i=1}^5U(1)_i$ as described in section \ref{geoengin}. If we instead consider $K'\equiv\prod_{i=1}^4U(1)_i$ gauge theory\footnote{Alternatively, if we had chosen to remove $U(1)_1$, then the geometry would have given rise to a $\overline{\mathbf{5}}$ of $SU_5$.}, then the manifold $\left(\mathbb{H}^6//K'\right)/U(1)_5$ is seven-dimensional, and is considered the fibration of ADE singularities over a base $Q$ that is coordinatized by the D/F-term of $U(1)_5.$

To see that the geometry corresponds to the resolution $SU_6\to SU_5$, let us write the D/F-term\footnote{Of course, the components of $\vec{\!\!~t}_5$ are related by an $SU_2$ symmetry and so there is nothing special about this choice of paramterization.} \begin{equation}\vec{\!\!~t}_5\equiv(\mathfrak{Re}(t),\mathfrak{Im}(t),r_t),\label{resolveT}\end{equation} where $t\in\mathbb{C}$ and $r_t\in\mathbb{R}$. Then using the same definitions of $x,y,$ and $z$ as in equation (\ref{xyz}), the definition of $\vec{\!\!~t}_5$ tells us that $z_5\overline{z}_5=z+t$ so that $x,y,$ and $z$ are related via $xy=z^5(z+t)$. Clearly this is the geometry we desired: at $\vec{\!\!~t}_5=0$ the fibre is $SU_6$ and at any $\vec{\!\!~t}_5\neq0$ it is $SU_5$.

\vspace{-0.35cm}\subsubsection{Engineering a $\mathbf{10}$ of $SU_5$\label{secso10}}\vspace{-0.35cm}
Recall that a $\mathbf{10}$ of $SU_5$ arises via the resolution $SO_{10}\to SU_{5}$. Therefore, we must deform an $SO_{10}$ singularity so as to give rise to $SU_5$ over a generic point on the base space $Q$.

\begin{figure*}[t]\includegraphics[scale=1.65]{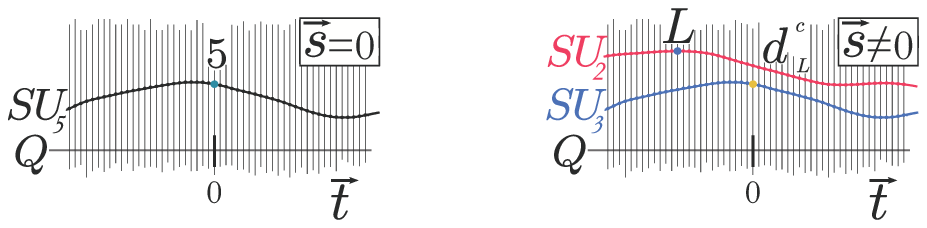}\caption{A representation of the fibre structure of the $G_2$-manifold found when unfolding a $\mathbf{5}$ of $SU_5$ into $SU_3\times SU_2$ by varying the deformation parameter $\vec{\!\!~s}$. When $\vec{\!\!~s}=0$, the geometry is precisely that which gives rise to a single massless $\mathbf{5}$ of $SU_5$. When $\vec{\!\!~s}\neq0$, however, the $SU_5$ singularity in each generic $K3$ fibre `splits apart' into two ADE-type singularities, giving rise $SU_3$ and $SU_2$ gauge theory; and what was once an islated conical singularity at $\vec{\!\!~t}=0$ splits into a pair of conical singularities, generating the matter content shown. \label{unfolding_a_5}}\end{figure*}

As we described above, the $SO_{10}$ singularity is described as a $\mathbb{Z}_2$-orbifold of an $SU_{10}$ singularity; and its resolutions are those of the $SU_{10}$ cover which are invariant under $\mathbb{Z}_2$. Using the now `canonical' labelling conventions to describe the $SU_{10}$ singularity, let us define $K'\equiv\prod_{i\neq5}U(1)_i$. Then the orbifolded hyper-K\"{a}hler quotient $\left[\left(\mathbb{H}^{10}//K'\right)/U(1)_5\right]/\mathbb{Z}_2$ is a seven-dimensional manifold which we will view as a fibration over a space parameterized by $\vec{\!\!~t}_5$, the D/F-term for $U(1)_5$, such that for $\vec{\!\!~t}_5\neq0$ the fibres are $SU_5$ and when $\vec{\!\!~t}_5=0$ the singularity is enhanced to $SO_{10}$. To demonstrate this claim, we will construct the complex structure of the fibres as we did for the $\mathbf{5}$ representation.

Again, we will use the same notation as above: \vspace{-0.15cm}\begin{equation}x\equiv\prod_{i=0}^{9}z_i,\quad y\equiv\prod_{i=0}^{9}\overline{z}_i,\quad z\equiv z_0\overline{z}_0.\vspace{-0.15cm}\end{equation} We will parameterize the D/F-term of $U(1)_5$ almost exactly as we did in equation (\ref{resolveT}), but with a factor of 2 inserted for future notational simplicity: \vspace{-0.15cm}\begin{equation}\vec{\!\!~t}_5\equiv2(\mathfrak{Re}(t),\mathfrak{Im}(t),r_t).\vspace{-0.15cm}\end{equation} Recall that the $\mathbb{Z}_2$ action defining the orbifold is generated by $S:(x,y,z)\mapsto(y,x,-z)$. Notice also that under $S$, $\vec{\!\!~t}_5\mapsto-\vec{\!\!~t}_5$.

Because the definition of the hyper-K\"{a}hler quotient giving the covering $SU_{10}$ singularity implies that the D/F-terms of $U(1)_{i\neq5}$ all vanish, we see that \vspace{-0.15cm}\[z_4\overline{z}_4=z_3\overline{z}_3=\ldots=z_0\overline{z}_0=z,\vspace{-0.25cm}\] and\vspace{-0.15cm} \[z_9\overline{z}_9=z_8\overline{z}_8=\ldots=z_5\overline{z}_5=z+2t.\]
Therefore, $x,y,$ and $z$ are related by \begin{equation}xy=z^5(z+2t)^5\quad\implies\quad xy=(\zeta^2-t^2)^5,\end{equation}
where we have introduced the new variable $\zeta\equiv z+t$. Notice that because both $z$ and $t$ are odd under $S$, so is $\zeta$. Of course, these variables are still not invariant under the action of $\mathbb{Z}_2$. This is largely remedied by defining the new variables\footnote{A keen observer will notice that $Y$ as defined here is {\it not} invariant under $\mathbb{Z}_2$. However, this does not pose any problems because equation (\ref{10ofsu5}), which relates $X,Y,$ and $Z$, is itself invariant under $\mathbb{Z}_2$.} \begin{align}X\equiv\frac{1}{2}\zeta^{-1}\left(x-y\right),&\quad Y\equiv\frac{1}{2}\zeta^{-2}\left(x+y+2t^5\right),\nonumber\\
\mathrm{and}\quad Z&\equiv\zeta^2,\end{align}
by analogy to equation (\ref{XYZ}) above. They are related by the identity \begin{equation}X^2=Y^2Z-Z^{-1}\left\{\left(Z-t^2\right)^5-t^{10}\right\}-2Yt^5.\label{10ofsu5}\end{equation}
This defining equation is invariant under the action of $\mathbb{Z}_2$ and so the surface of solutions to this equation in $\mathbb{C}^3$ is isomorphic to the fibre over $\vec{\!\!~t}_5$ (for vanishing $r_t$). Equation (\ref{10ofsu5}) should be recognized as precisely the deformation $SO_{10}\to SU_5$ giving rise to a massless $\mathbf{10}$ of $SU_5$ (see e.g. \cite{Katz:1996xe} for details).

\section{Unfolding $SU_5$ Representations into the Standard Model\label{unfold}}
In our discussions above, only one D/F-term of the linear sigma model was considered as potentially acquire a non-zero value; and this D/F-term was given a special interpretation as parameterizing the singular, four-dimensional fibres over a three dimensional base $Q$. As described analogously in \cite{Bourjaily:2007vw,Bourjaily:2007vx}, if however one of these parameters were allowed to be non-vanishing {\it independent of the location on $Q$}, the type of singularity over the whole of $Q$ would reduce in rank---corresponding to `unfolding' the manifold into one with less symmetry. When isolated enhanced singularities are present, they also reduce in rank but remain locations of enhanced singularity relative to the singularity of the generic fibre over $Q$. In general, the conical, enhanced singularities giving rise to charged matter will unfold into a collection of disparate singularities giving rise to the representations which would have resulted from ordinary symmetry breaking.

In this section we will make this idea explicit by unfolding the singularities giving rise to $SU_5$ matter representations into their corresponding representations of $SU_3\times SU_2$.

\vspace{-0.35cm}\subsection{Unfolding a $\mathbf{5}$ of $SU_5$ into $(\overline{\mathbf{3}},\mathbf{1})\oplus(\mathbf{1},\mathbf{2})$ of $SU_3\times SU_2$}\vspace{-0.35cm}
Let us describe how the conical singularity engineered in section \ref{5ofsu5} can be unfolded into the Standard Model. To do this, we must set the D/F-term of one (or more) of the other $U(1)_i$'s to some constant value, independent of the location $\vec{\!\!~t}_5$ on the base $Q$, so that the generic fibre over $\vec{\!\!~t}_5$ is $SU_3\times SU_2$. Thinking about the linear sigma model as a quiver theory, it is easy to see which $U(1)$ should be resolved. Indeed, the resolution of either $U(1)_2$ or $U(1)_3$ would result in $SU_3\times SU_2$ singularities in the generic fibre; a flip of a coin chose $U(1)_2$ for us.

Let us define $K'\equiv U(1)_1\times U(1)_3\times U(1)_4$. The vacuum manifold of the linear sigma model of the $6$ hypermultiplets $\Phi_{i}$---denoted $(\mathbb{H}^6//K')$---is then a twelve-dimensional hyper-K\"{a}hler manifold. By quotienting by $U(1)_2$ and $U(1)_5$, we then obtain the ten-dimensional hyper-K\"{a}hler quotient $(\mathbb{H}^6//K')/(U(1)_2\times U(1)_5)$. The D/F-terms of the two `resolved' $U(1)$'s parameterize a six-dimensional base over which are fibred ADE singularities.

\begin{figure*}[t]\includegraphics[scale=1.65]{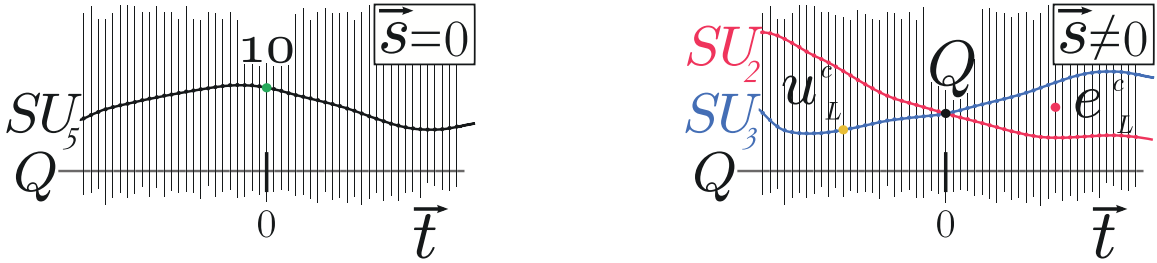}\caption{A representation of the fibre structure of the $G_2$-manifold found when unfolding a $\mathbf{10}$ of $SU_5$ into $SU_3\times SU_2$ by varying the deformation parameter $\vec{\!\!~s}$. When $\vec{\!\!~s}=0$, the geometry is precisely that which gives rise to a single massless $\mathbf{10}$ of $SU_5$. When $\vec{\!\!~s}\neq0$, however, the $SU_5$ singularity in each $K3$ fibre `splits apart' into two ADE-type singularities giving rise $SU_3$ and $SU_2$ gauge theory; and what was once an isolated conical singularity at $\vec{\!\!~t}=0$ splits into three distinct conical singularities, generating the matter content shown.\label{unfolding_a_10}}\end{figure*}

We will continue to use $\vec{\!\!~t}_5\equiv\vec{\!\!~t}\equiv(\mathfrak{Re}(t),\mathfrak{Im}(t),r_t)$ to parameterize the fibres over $Q$, while we will use \mbox{$\vec{\!\!~t}_2\equiv\vec{\!\!~s}\equiv(\mathfrak{Re}(s),\mathfrak{Im}(s),r_s)$} to parameterize deformations of all the fibres over $Q$ independent of $\vec{\!\!~t}$.

Using the variables $x,y,$ and $z$ as defined in equation (\ref{xyz}), we see that they are now related by \begin{equation}xy=z^2(z+s)^3(z+t+s).\end{equation} This allows us to conclude directly that the generic fibre over $Q$ has both an $SU_3$ and an $SU_2$ singularity, giving rise to $SU_3\times SU_2$ gauge theory. Furthermore, we see that when $\vec{\!\!~t}=0$, the singularity is enhanced to $SU_4\times SU_2$ giving rise to matter in the $(\overline{\mathbf{3}},\mathbf{1})$ representation of $SU_3\times SU_2$; and at $\vec{\!\!~t}=-\vec{\!\!~s}$ the singularity is enhanced to $SU_3\times SU_3$ giving rise to matter in the $(\mathbf{1},\mathbf{2})$ representation. A cartoon of the fibration is shown in Figure \ref{unfolding_a_5}.

Therefore, we have explicitly shown that the enhanced, conical singularity of a $G_2$-manifold giving rise to a massless $\mathbf{5}$ of $SU_5$ can be smoothly deformed into two enhanced conical singularities, giving rise to matter in the $(\overline{\mathbf{3}},\mathbf{1})$ and $(\mathbf{1},\mathbf{2})$ representations of $SU_3\times SU_2$ gauge theory. 

\subsection{Unfolding a $\mathbf{10}$ of $SU_5$ into $(\mathbf{3},\mathbf{2})\oplus(\overline{\mathbf{3}},\mathbf{1})\oplus(\mathbf{1},\mathbf{1})$ of $SU_3\times SU_2$}\vspace{-0.35cm}
Recall that because the $SO_{10}$ singularity was constructed as a $\mathbb{Z}_2$-orbifold of an $SU_{10}$ singularity, we should describe the resolutions of $SO_{10}$ in terms of the resolutions of its cover-space $SU_{10}$ which are invariant under the action of $\mathbb{Z}_2$. This makes it slightly harder to guess which $U(1)$'s to resolve. Indeed, in the spirit of the multiple unfoldings described in \cite{Berglund:2002hw}, we will find that the resolution must be a linear combination of D/F-terms from two different $U(1)$'s. Rather than motivating the resolution as we did above, we will simply state the answer and verify that it does indeed produce the desired structure.

Using the language and notation of section \ref{secso10}, let \[K'\equiv U(1)_1\times U(1)_2\times U(1)_4\times U(1)_6\times U(1)_7\times U(1)_9.\] Then the (orbifolded) hyper-K\"{a}hler quotient \[\left[\left(\mathbb{H}^{10}//K'\right)/\left(U(1)_3\times U(1)_5\times U(1)_8\right)\right]/\mathbb{Z}_2,\] is a thirteen-dimensional manifold where the D/F-terms of the three resolved $U(1)$'s parameterize a family of ADE singularities. As in section \ref{secso10} we will use $\vec{\!\!~t}_5\equiv\vec{\!\!~t}\equiv2(\mathfrak{Re}(t),\mathfrak{Im}(t),r_t)$ to parameterize the fibres over the three-dimensional base $Q$. The D/F-terms for $U(1)_3$ and $U(1)_8$ will be related to one another by \[\vec{\!\!~t}_3=-\vec{\!\!~t}_8\equiv\vec{\!\!~s}\equiv2(\mathfrak{Re}(s),\mathfrak{Im}(s),r_s).\]
We will show that if $\vec{\!\!~s}$ is takes on a non-zero value---independent of the location on $Q$---then the manifold will be found to give rise to \mbox{$SU_3\times SU_2$} gauge theory with the singularities supporting matter in the $(\mathbf{3},\mathbf{2})$, $(\overline{\mathbf{3}},\mathbf{1})$ and $(\mathbf{1},\mathbf{1})$ representations of $SU_3\times SU_2$.

Using the now familiar definitions of $x$ and $y$, and continuing to let $z\equiv z_0\overline{z}_0$, we see that here they are related by\vspace{-0.25cm} \begin{align}xy&=z^3(z+2s)^2(z+2(s+t))^3(z+2t)^2,\nonumber\\
\implies\quad xy&=\left(\zeta^2-(t+s)^2\right)^3\left(\zeta^2-(t-s)^2\right)^2,\nonumber\end{align}
where we have introduced the variable $\zeta\equiv z+t+s$. Notice that $\zeta$ is odd under the action of $S$. Proceeding as we did in section \ref{secso10}, we introduce the variables \vspace{-0.15cm} \begin{align}X&\equiv\frac{1}{2}\zeta^{-1}\left(x-y\right),\quad Z\equiv\zeta^2,\nonumber\\
\mathrm{and}\quad Y&\equiv\frac{1}{2}\zeta^{-2}\left(x+y+2(t-s)^2(t+s)^3\right).\end{align}
And we find that these new variables are related by, \begin{widetext}\vspace{-0.45cm}\begin{equation}X^2=Y^2Z-Z^{-1}\left\{\left(Z+(t+s)^2\right)^3\left(Z-(t-s)^2\right)^2-\left(t+s\right)^6\left(t-s\right)^4\right\}-2Y\left(t+s\right)^3\left(t-s\right)^2.\label{so10cstruct}\end{equation}\end{widetext}
This equation is invariant under the action of $\mathbb{Z}_2$ and so therefore describes the fibre over a point $\vec{\!\!~t}$ (for vanishing $r_t$). This is precisely the same complex structure encountered in \cite{Bourjaily:2007vw}. Therefore, we see that for fixed $\vec{\!\!~s}\neq0$, the generic fibre over $\vec{\!\!~t}$ is simply $SU_3\times SU_2$, and there are three isolated places in $Q$ where the singularity is enhanced: at $\vec{\!\!~t}=-\vec{\!\!~s}, \vec{\!\!~t}=0,$ and $\vec{\!\!~t}=\vec{\!\!~s}$ there are enhancements of the singularities giving rise to matter in $(\overline{\mathbf{3}},\mathbf{1})$, $(\mathbf{3},\mathbf{2})$, and $(\mathbf{1},\mathbf{1})$ representations of $SU_3\times SU_2$, respectively. The fibrtion of the unfolded geometry is illustrated in Figure \ref{unfolding_a_10}.

\vspace{-0.35cm}\section{Discussion}\vspace{-0.35cm}
In this paper we have shown that the basic ideas of \cite{Bourjaily:2007vw,Bourjaily:2007vx} have a concrete realization in M-theory. This is in some sense a proof of concept that also in M-theory, geometrically-engineered grand unified models can be unfolded into ones with less gauge symmetry. And although we have presented only the simplest example of interest, we have every reason to suspect that all the structure described in \cite{Bourjaily:2007vx} giving rise to three families can also be realized in M-theory.

We should clarify that here, like in \cite{Bourjaily:2007vw,Bourjaily:2007vx}, there is no clear sense in which the traditional notion of `symmetry breaking' is taking place: because we do not yet understand the dynamics which control and set the values of the parameters unfolding the geometry, it isn't at all obvious that `Higgsing' makes dynamical sense here. From a rather conservative point of view, our work merely provides a motivation for a possible local origin of the three families of the Standard Model (extended similarly to $E_6$ models). One potentially interesting feature of this motivating picture is that the relative locations of the diverse conical singularities can be related to one another by the parameters which deform the more unified geometry. And because the geometric structures giving rise to the Standard Model matter content are all local in nature, this may allow string phenomenologists to separate the discussion of global stability from one of phenomenology. The general philosophy behind our local, geometric origin of the three families is quite similar to that described in \cite{Verlinde:2005jr}.

A strength of this approach is that one can motivate and in principle describe the local geometries of at least one class of $G_2$-manifolds that have all the structure necessary to have $SU_3\times SU_2$ gauge theory with three families at low energy. Being a local description of the compactification manifold, however, this picture allows for a number of potential criticisms. Recall that we have only described a single coordinate patch along the base space $Q$, which we have taken to be $\mathbb{R}^3$. This surely cannot be the whole story: if $Q$ were $\mathbb{R}^3$, not only would the resulting theory have no gravity ($m_{\mathrm{Pl}}\to\infty$ by non-compactness), but it would effectively have no gauge theory either ($g_{\mathrm{YM}}\to0$): the low-energy gauge coupling is set by $1/\mathrm{Vol}(Q)$.

So clearly the local patch of $Q$ which we describe must be glued into a compact three-fold. And while we are able to motivate much of the low energy physics---three families, an interesting superpotential, etc.---without knowledge of the global topology of $Q$, we are not able to ignore its global structure entirely. For example, unless we desire chiral fields charged under the adjoint of the gauge group, the fundamental group of $Q$ must be finite (see e.g. \cite{Acharya:2004qe}); also, the local patch should be completed in such a way that no additional, unwanted charged matter arises. 

Whether or not these non-compact, local $G_2$ constructions can be compactified is a very difficult mathematical question that may not find an answer for a long time\footnote{Indeed, there is not even a single example of a compact $G_2$-manifold with a conical singularity. But there are strong reasons to suspect they are not at all uncommon: because M-theory on a $K3$-fibred $G_2$-manifold is dual to the heterotic string on a $T^3$-fibred Calabi-Yau three-fold, for example, the wide variety of heterotic constructions suggests that conical singularities in $G_2$-manifolds are not uncommon in moduli space (see e.g. \cite{Acharya:2004qe}).}. In principle, compactification should break the seemingly continuous range of possible values for the deformation moduli into a discrete landscape; how large that landscape is, and whether or not it admits models consistent with experiment are of course very important questions that we are not able to answer yet today. 

But one of the assets of our approach is the ability to describe a great deal of low-energy phenomenology before these mathematical challenges are overcome---that is, without full knowledge of the Planck-scale physics and global topology. It is important then to ask under what conditions this separation of the problem is self-consistent. This should be the case if all the singularities giving rise to matter at low-energy are separated by distances much smaller than the size of $Q$. Such a hierarchy of scales could be natural for example in the context of a warped compactification geometry. 

In this paper we have shown that the idea of unfolding geometrically engineered grand unified models as described in \cite{Bourjaily:2007vw} is also possible in M-theory. This suggests that if we continued our discussion to more unified initial geometries, we would also find the three families of the Standard Model unfolding naturally out of the an initial $E_8\to E_6\times SU_2$ resolution. This could possibly provide a compelling explanation for the origin of three families in the Standard Model. Furthermore, because the unfolded local geometry can in principle be completely specified, this approach is a modest step toward concrete phenomenological analyses of at least one class of models in M-theory.

\vspace{-0.5cm}\section{Acknowledgements}\vspace{-0.35cm}
The author appreciates many helpful discussions with and suggestions from Herman Verlinde, Sergei Gukov, Gordon Kane, Edward Witten, Paul Langacker, Bobby Acharya, Per Berglund, Liantao Wang, Dmitry Malyshev, Matthew Buican, Piyush Kumar, and Konstantin Bobkov.

This research was supported in part by a Graduate Research Fellowship from the National Science Foundation.


\begin{thebibliography}{10}

\bibitem{Bourjaily:2007vw}
J.~L. Bourjaily, ``Multiple {U}nfoldings of {O}rbifold {S}ingularities:
  {E}ngineering {G}eometric {A}nalogies to {U}nification,'' 2007,
  arXiv:0704.0444 [hep-th].

\bibitem{Bourjaily:2007vx}
J.~L. Bourjaily, ``Geometrically {E}ngineering the {S}tandard {M}odel:
  {L}ocally {U}nfolding {T}hree {F}amilies out of ${E}_8$,'' 2007,
  arXiv:0704.0445 [hep-th].

\bibitem{Horava:1996ma}
P.~Horava and E.~Witten, ``Eleven-{D}imensional {S}upergravity on a {M}anifold
  with {B}oundary,'' {\em Nucl. Phys.}, vol.~B475, pp.~94--114, 1996,
  hep-th/9603142.

\bibitem{Atiyah:2001qf}
M.~Atiyah and E.~Witten, ``M-theory {D}ynamics on a {M}anifold of ${G}_2$
  {H}olonomy,'' {\em Adv. Theor. Math. Phys.}, vol.~6, pp.~1--106, 2003,
  hep-th/0107177.

\bibitem{Witten:2001uq}
E.~Witten, ``Anomaly {C}ancellation on ${G}_2$ {M}anifolds,'' 2001,
  hep-th/0108165.

\bibitem{Acharya:2001gy}
B.~Acharya and E.~Witten, ``Chiral {F}ermions from {M}anifolds of ${G}_2$
  {H}olonomy,'' 2001, hep-th/0109152.

\bibitem{Acharya:2004qe}
B.~S. Acharya and S.~Gukov, ``M-theory and {S}ingularities of {E}xceptional
  {H}olonomy {M}anifolds,'' {\em Phys. Rept.}, vol.~392, pp.~121--189, 2004,
  hep-th/0409191.

\bibitem{Katz:1996xe}
S.~Katz and C.~Vafa, ``Matter from {G}eometry,'' {\em Nucl. Phys.}, vol.~B497,
  pp.~146--154, 1997, hep-th/9606086.

\bibitem{Berglund:2002hw}
P.~Berglund and A.~Brandhuber, ``Matter from ${G}_2$ {M}anifolds,'' {\em Nucl.
  Phys.}, vol.~B641, pp.~351--375, 2002, hep-th/0205184.

\bibitem{Verlinde:2005jr}
H.~Verlinde and M.~Wijnholt, ``Building the {S}tandard {M}odel on a
  {D}3-{B}rane,'' {\em JHEP}, vol.~01, p.~106, 2007, hep-th/0508089.

\bibitem{Katz:1992ab}
S.~Katz and D.~Morrison, ``Gorenstein {T}hreefold {S}ingularities with {S}mall
  {R}esolutions via {I}nvariant {T}heory for {W}eyl {G}roups,'' {\em J.
  Algebraic Geometry}, vol.~1, pp.~449--530, 1992.

\bibitem{Kronheimer:1989zs}
P.~B. Kronheimer, ``The {C}onstruction of {A}{L}{E} {S}paces as hyper-{K}\"{a}hler
  {Q}uotients,'' {\em J. Diff. Geom.}, vol.~29, pp.~665--683, 1989.

\bibitem{Kronheimer:1989pu}
P.~B. Kronheimer, ``A {T}orelli {T}ype {T}heorem for {G}ravitational
  {I}nstantons,'' {\em J. Diff. Geom.}, vol.~29, pp.~685--697, 1989.

\end{thebibliography}

\end{document}